   \documentclass[prl,aps,twocolumn,showpacs]{revtex4}
\usepackage[dvips]{graphicx}
\usepackage{dcolumn}%
\usepackage{amsmath}%
\usepackage{amsfonts}%
\usepackage{amssymb}
\providecommand{\U}[1]{\protect\rule{.1in}{.1in}}
\newcommand{\be}{\begin{equation}}
\newcommand{\ee}{\end{equation}}
\newcommand{\bea}{\begin{eqnarray}}
\newcommand{\eea}{\end{eqnarray}}

\begin{document}
%%\large

\title{Electron transport through asymmetric ferroelectric tunnel junctions:\\ current-voltage characteristics}

\author{ Natalya A. Zimbovskaya}

\affiliation{Department of Physics and  Electronics, University of Puerto Rico, 100 CUH Station, Humacao, PR 00791, and Institute for Functional Nanomaterials, University of Puerto Rico, San Juan, PR 00931}

\begin{abstract} 
 We have carried out calculations of current-voltage characteristics for the electron tunnel current through a junction with a thin insulating ferroelectric barrier assuming that interface transmissions for the left and right interfaces noticeably differ due to dissimilarity of the interfaces.  Obtained conductance vs voltage and current vs voltage curves exhibit well distinguishable asymmetric hysteresis. We show that the asymmetry in the hysteretic effects could originate from the asymmetric bias voltage profile inside the junction. In particular, we analyze the hysteresis asymmetries occurring when the bias voltage distribution is low-sensitive to the spontaneous polarization reversal.
 \end{abstract}

\pacs{73.23.-b, 85.50.-n}
\date{\today}
\maketitle

Ferroelectric materials attract significant interest of the research community due to their potential usefulness in various technological applications \cite{1}. In the past decade it was repeatedly shown in the experiments that ferroelectricity could be maintained in films of some  perovskite oxides with thickness of the order of a few nanometers \cite{2,3,4,5,6}. These experimental observations were corroborated by first-principles computations predicting the critical film thickness for ferroelectricity to persist in perovskite films to be as thin as a few lattice parameters \cite{7,8,9,10}.   Thus, both experimental and theoretical results indicate that ultrathin ferroelectric   films exist and may be used as barriers in ferroelectric tunnel junctions (FTJ). A ferroelectric tunnel junction consists of two conducting electrodes (leads) separated by a nanometer-thick ferroelectric film. Electrons can tunnel across the  film thus maintaining a nonzero conductance of the junction. Specific nature of the ferroelectric barriers add new functional properties to FTJs \cite{11}
 Such junctions could serve as essential components in manifold nanodevices such as binary data storage memories and switches  \cite{12}.  

The most important feature of ferroelectrics is the presence of spontaneous polarization which could be reversed by an applied electric field. The polarization switching inherent to ferroelectric materials alters the sign of polarization charges at the electrode/barrier interfaces thus changing the electrostatic potential profile inside the junction. Also, the polarization switching shifts ions in ferroelectric unit cells and changes lattice strains in the film affecting the electron band structure of the junction. As a result, transport characteristics of a FTJ may be significantly modified. The effect of the polarization reversal on FTJ conduction was predicted in earlier works  (see e.g. Refs. \cite{13,14,15}) and observed in experiments on $Pt/Pb(Zr_{0.52}Ti_{0.48})O_3/SrRuO_3 $ FTJ \cite{15}, on junctions with $La_{0.1} Bi_{0.9} MnO_3$ tunnel barrier \cite{16} and on $ Pt/SrTiO_3/Pt $ junctions \cite{17}.

Currently, theoretical and computational efforts are mostly concentrated on studies of zero bias conductance of FTJs. However, the theoretical analysis of current-bias voltage $(I-V)$ relationships is also important for it provides more thorough characterization of the electron transport through FTJs. As shown in Ref. \cite{13}, the polarization reversal in the ferroelectric films causes a hysteretic behavior of both current-voltage and conductance-voltage characteristics. Especially interesting hysteretic effects may occur in asymmetric FTJs where the interfaces are dissimilar. The dissimilarity could occur when the leads are made out of different metals \cite{14,18}. Also, it could originate from the atomic and electronic structure of the junction \cite{19}.  When interfaces differ, the effects of the polarization reversal also may differ depending on the specific properties of the interfaces. This may result in asymmetry of tunnel electroresistance (TER) versus bias voltage curves and of hysteresis loops in current-voltage characteristics. In strongly asymmetric FTJs both TER and hysteresis could be well pronounced for a certain bias voltage polarity and significantly reduced on the polarity reversal. The asymmetric TER and hysteretic effects were recently reported for asymmetric junctions including $Ba Ti O_3 $ \cite{20,21} and $ Pb(Zr_{0.2}Ti_{0.8})O_3$  \cite{22} ferroelectric barriers.

 In the present work we theoretically analyze transport characteristics of an asymmetric FTJ. 
We simulate the FTJ by a ferroelectric film of thickness $ d $ sandwiched between two semi-infinite metal electrodes. We assume that the film is uniformly polarized in the direction perpendicular to the electrode planes, which is the direction of $``z"$ axis within the chosen coordinate system. At small values of the applied bias voltage the conductance of a tunnel junction per area $ A$ could be described by the following expression \cite{23}:
  \be
  \frac{G}{A} = \frac{2e^2}{h} \int \frac{d^2 \bf k_{||}}{(2\pi)^2} T(E,\bf k_{||})                \label{1}                                      \ee  
  where $\bf k_{||} $ is the projection of the tunneling electron wave vector to the electrode's plane, and $ E $ is the tunnel energy, which takes on values close to the Fermi energy $E_F.$ The transmission function $ T(E,\ \bf k_{||})$ may be factorized and presented in the form \cite{24}:
  \be
  T(E,{\bf k_{||}}) = t_L (E,{\bf k_{||}}) \exp \big \{ - 2 \kappa (E,{\bf k_{||}}) d \big\}t_R (E,{\bf k_{||}})   \label{2} \ee
  Here,  $ t_L (E,{\bf k_{||}}) $ and $ t_R (E,{\bf k_{||}})$ are the interface transmission functions for the left and right electrodes, respectively, and
$\kappa (E,{\bf k_{||}})$ is determined by the height and shape  of the potential barrier in the tunnel junction and by the energy-momentum relation of a tunneling electron. Generally speaking, it also depends on the bias voltage but one may disregard this at small values of $V.$ Employing a simple form for the energy-momentum relation:
   \be  
  E = \frac{\hbar^2k_{||}^2}{2m} + \frac{\hbar^2 k_z^2}{2m}   \label{3}
     \ee 
  where $ m$ is the effective mass of the tunneling electron, we may approximate:
    \be
  \kappa (E,{\bf k_{||}}) = \frac{1}{d} \int_0^d \sqrt{\frac{2m\Phi(z)}{\hbar^2} + k_{||}^2}\, dz  \label{4} 
  \ee
   Here, $ \Phi(z) $ is the overall potential profile seen by the tunneling electrons.  At zero bias voltage the potential $ \Phi(z)$ is a superposition of the potential, which determines the positions of conduction bands in metal electrodes with respect to the Fermi energy, and the barrier potential occurring due to the presence of the insulating ferroelectric film. Also, the overall potential includes a contribution created by surface charges on the surfaces of the ferroelectric film and screening charges appearing on the electrode surfaces. The simplest approximation for $\Phi (z) $  appropriate for a symmetric FTJ is a trapezoid barrier with a thickness $ d $ and heights $ \Phi (0) = U + \varphi(0),\ \Phi(d) = U + \varphi(d) $ where $ \varphi(z) $ is the potential induced by the surface charges. The relative heights of the barrier at $z=0 $ and $ z = d $ are determined by the potential difference $ \varphi(0) - \varphi(d).$ The difference changes its sign when the polarization in the film reverses causing the hysteresis in the conduction through the FTJ to occur. In asymmetric FTJs  the change in the complex band structure induced by the left-right asymmetry of the ferroelectric displacements in the barrier may noticeably change the average barrier heights $ U $ for two polarization states, as was theoretically shown for a $SrRuO_3/BaTiO_3/SrRuO_3 $  junction basing on first principles computations \cite{19}. The difference in the barrier heights for different polarization directions may be of the same order or greater than that induced by the potential $\varphi (z).$ 
 In  the following transport calculations we take into account both the effect of depolarizing field corresponding to the potential $ \varphi (z) $ and variations in the barrier height appearing due to the junction asymmetry. However, to avoid extra complications in further computations we do not include in our model the effect of an intrinsic electric field throughout the barrier, which appears due to the difference in the electrostatic dipoles at the interfaces, as well as the strain effects due to piezoelectricity.

To estimate the current through the junction we use the expression:
   \be  \frac{J}{A} = \frac{2e}{h} \int dE (f_L - f_R) \int \frac{d^2 \bf k_{||}}{(2\pi)^2} T(E,\bf k_{||}).        \label{5}
  \ee
   Here, the transmission function $ T(E,\bf k_{||})$ is given by Eq. (\ref{2}) and $f_{L,R}$ are Fermi distribution functions with chemical potentials $\mu_{L,R}.$ The chemical potentials are shifted with respect to $ E_F$ when a nonzero bias voltage $ V $ is applied across the junction, namely:
  \be   \mu_L = E_F + \eta eV; \qquad \mu_R = E_F - (1- \eta) eV.   \label{6}
     \ee
  The division parameter $ \eta $ shows how the voltage $ V $ is distributed in the junction. The expression (\ref{5}) resembles the well known Landauer formula for tunnel  current through molecular junctions and/or quantum dots coupled to the conducting leads \cite{25}. It could be derived using surface Green's functions formalism presented in the earlier work \cite{26}.  
 We have grounds to belive that $\eta $ takes on  values depending on the relation of the interface transmissions $ t_L $ and $ t_R.$ Calculating the current density we hypothesize that $ \eta = t_R/(t_L + t_R), $ which resembles the expression for the division parameter commonly used in the theory of electron tunnel transport through molecular bridges and quantum dots. In that theory $\eta $ is described by the similar expression where the parameters determining the coupling of a molecule/quantum dot to the left and right leads take on the parts of the interface transmissions. 

Regardless of symmetry/asymmetry of a FTJ, the transmission coefficients $t_L $ and $t_R $ differ, and the difference in their values may be significant. For instance, it was reported that in a $ Pt/BaTiO_3/Pt$ junction one of the transmission coefficients was approximately three times greater than another one \cite{10}. One may expect even greater difference in the interface transmission for the FTJ studied by Maksymovich et al \cite{22}, where rectifying current-voltage characteristics were reported. One reason for the disparity in the transmission functions is the trapezoid-like profile of the potential barrier. The barrier height decreases along the polarization direction, therefore $t_R $ should exceed $ t_L $ when the polarization points to the right and vice versa. In symmetrical FTJs with identical interfaces one may expect the transmission coefficients simply interchange their values on the polarization reversal:
    \be
 t_L^{\rightarrow} = t_R^{\leftarrow}; \qquad
 t_R^{\rightarrow} = t_L^{\leftarrow}                       \label{7} 
  \ee
 where an arrow indicates the polarization direction. In such a case the bias voltage division parameter $ \eta $ in the Eq. (\ref{6}) is to be replaced by $ 1 - \eta $ when the polarization direction is reversed. Accordingly, the current-voltage curves associated with different polarization states (although asymmetric by themselves due to the asymmetry in the bias voltage distribution across the junction) are arranged in the $ V-I $ plane in such a way that hysteretic features remain symmetrical. The TER versus voltage curve must be symmetrical upon bias voltage reversal, as well.

In asymmetric FTJs the relationship between the transmissions $ t_L $ and $ t_R $ for different polarization directions is more complicated. Due to an asymmetric deformation of the ferroelectric potential profile the relations (\ref{7}) do not hold any more. Besides, dissimilarity in the interfaces may affect the transmissions due to the reasons unrelated to ferroelectric properties of the film linking the electrodes. Even within the simple one-dimensional model of a FTJ employed in the present work, one must take into account specific features of the contact between the electrodes and the film, which could significantly differ on the left anf right sides of the junction. For instance, in the case when the electrodes are made out of different metals,  this may significantly  influence the interface transmissions. Also, FTJs studied in the recent works \cite{20,22} consisted of an epitaxial ferroelectric film grown on the top of conducting electrode and a sharp metal tip serving as another electrode. We may expect a nonepitaxial contact between the tip and the film to notably modify the corresponding interface transmission for both polarization directions. 
  Some other effects, which could modify interface transmissions in asymmetric FTJs (and, in consequence, the bias voltage distribution across the junctions) are discussed elsewhere (see e.g. Ref. \cite{22} and references therein).
 So, the  bias voltage distribution across an asymmetric FTJ for different polarization orientations could hardly be predicted basing on some general considerations. For each particular FTJ the relation of the interface transmissions (which greatly affects the bias voltage distribution) must be separately established, and the results may significantly vary depending on the FTJ characteristics. For instance, we remark that in the asymmetric FTJ whose transport properties were studied by Gruverman et al \cite{20}, $ t_L < t_R $ for both polarization directions. This follows from the reported shape of the $ I-V $ characteristics. 

In summary, various asymmetric features in the hysteretic behavior of the tunnel current and conductance in asymmetrical FTJs originate from asymmetries in the tunnel barrier profile for different polarization directions and from specifics of the bias voltage distribution (determined by the relationship between the interface transmissions). The relative effects of these two factors could vary but neither one should be disregarded on general grounds.
   Here, we concentrate on the analysis of transport properties of a strongly asymmetric FTJ where the bias voltage distribution exhibits a relatively low sensibility to the polarization reversal. Accordingly, we accept that $ t_L/t_R $ remains the same regardless of the polarization direction.
   
\begin{figure}[t]  %%%fig. 1
\begin{center}
\includegraphics[width=5cm,height=9cm,angle=-90]{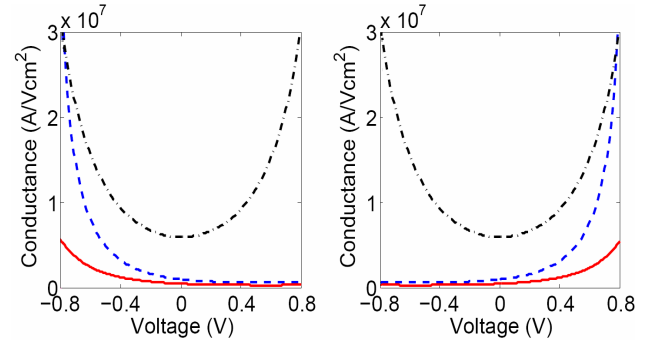}

\end{center}
\caption{(Color online)  Conductance $ \frac{G}{A} = \frac{1}{A} \frac{dI}{dV}$ of the model asymmetric FTJ vs bias voltage for the paraelectric (dash-dotted line) and ferroelectric (solid and dashed lines) states of the film. The curves are plotted using Eqs. (\ref{1})-(\ref{6}) at $T=30 K,\  d = 1.8 nm,\ U= 0.6 eV $ (low resistance ferroelectric state), $U = 0.8 eV$ (high resistance ferroelectric state), for $ t_L = 3t_R$ (left panel) and $ t_R = 3 t_L $ (right panel).
}%
\label{rateI}%
\end{figure}

Computing the conductance and current for our model FTJ we used the results reported in Ref.  \cite{10} for $Pt/BaTiO_3/Pt $ junctions. So, we assumed that $ d = 1.8 nm, $ and zero bias conductances per unit cell area take on values $ G/A \approx 17.0 \times 10^{-5} e^2/h $ (for the paraelectric state of the film) and $ G/A \approx 2.9\times 10^{-5} e^2/h $ (for its low resistance ferroelectric state), respectively \cite{27}. We did carry out calculations for $ t_R/t_L = 3\ (\eta =0.75)$ and $ t_L/t_R = 3\ (\eta =0.25)$ for ferroelectric states of the film. For the film in the paraelectric state we assumed $t_L = t_R.$
 Using these values we estimated the interface transmissions for paraelectric and ferroelectric barriers in the junction.    As shown in the Fig. 1, the conductance through the junction with the ferroelectric barrier should exhibit rather well distinguishable hysteresis. The hysteresis is asymmetric with respect to $ V = 0, $ and we observe the significant asymmetry  in the TER. 
 At $t_L = 3t_R $ the TER is small at positive bias voltage but it significantly increases upon the reversal of the bias voltage polarity, as shown in the left panel of the figure. When $ t_R = 3t_L $ the TER is much better pronounced at positive bias voltage (see the right panel of the Fig. 1). Basing on the experimental results of Garcia et al \cite{21} for thin $ BaTiO_3 $ films, we may roughly estimate the coercive voltage $V_c $ for our model FTJ as: $V_c \sim 1.0-1.5V.$ Therefore, the difference in the TER values at $ V = \pm V_c  , $ which occurs due to the specifics of the bias voltage profile across the junction may reach values of the order of $10^2- 10^3. $         The current-voltage characteristics are presented in the Fig. 2. 
 We see asymmetric hysteretic behavior of the tunnel current. As well as for conductance, the asymmetry in the hysteresis loops to a considerable degree originates from the bias voltage profile in the junction. We remark that the $ I-V $ curves in the right panel of the Fig. 2 resemble those recently reported in the work \cite{20}.

\begin{figure}[t]  %%%fig. 2
\begin{center}
\includegraphics[width=5cm,height=9cm,angle=-90]{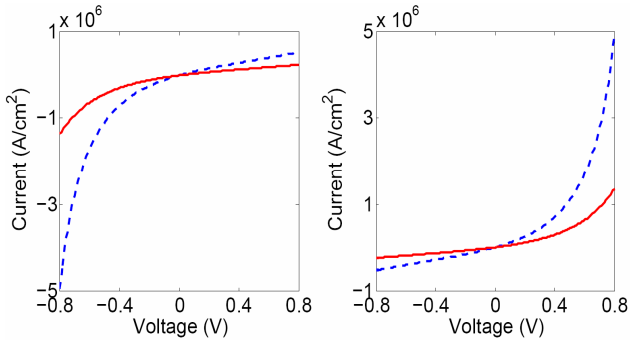}
\end{center}
\caption{(Color online)  Current-voltage characteristics for the model ferroelectric tunnel junction. The curves are plotted using Eqs. (\ref{1})-(\ref{6})  The  parameters take on the same values as in the Fig. 1.}%
\label{rateI}%
\end{figure}

In conclusion, we have demonstrated that current-voltage and conductance-voltage characteristics of an asymmetric FTJ may exhibit a pronounced hysteretic behavior of a special kind whose distinctive feature is a well pronounced asymmetry in the hysteresis loops for different bias voltage polarities. The asymmetry in the hysteresis largely originates from the asymmetric bias voltage division in the junction. In strongly asymmetryc junctions the dissimilarity of the interfaces may affect interface transmissions so much that the effects of the polarization reversal may be surpassed. In this case, the relationship between the transmissions is low sensitive to the polarization revesal, and the bias voltage distribution across the junction rather weakly depends on the polarization direction. 
Our results give grounds to expect that the bias voltage reversal in the considered strongly asymmetric FTJs should bring significant changes to their transport characteristics, and this could be used in designing of nanodevices.

{\it Acknowledgement:}
Author thank J. P. Velev for helpful discussion and  G. M. Zimbovsky for help in manuscript.


\begin{thebibliography}{99}
\bibitem{27} It was shown in Ref. \cite{10} that the $ BaTiO_3 $ film with the thickness of $1.8 nm $ could exist in both paraelectric  and ferroelectric states.
\end{thebibliography}
\end{document}